\documentclass{iopart}

\usepackage{graphicx}

\begin{document}
\input epsf

\title {Model of the mixed state of type-II superconductors in high 
magnetic fields}

\author {I. L. Landau\dag \ddag  and H. R. Ott\dag}
\address{\dag\ Laboratorium f\"ur Festk\"orperphysik, ETH H\"onggerberg, 
CH-8093 Z\"urich, Switzerland}
\address{\ddag\ Kapitza Institute for Physical Problems, 117334 Moscow, 
Russia}

\date{\today}

\begin{abstract}
In superconductors with large values of the Ginzburg-Landau parameter 
$\kappa$, exposed to magnetic fields close to the upper critical field 
$H_{c2}$, the magnetic field is practically homogeneous across the sample 
and the density of supercurrents is negligibly small. In this case, there 
is no obvious reason for the formation of Abrikosov vortices, characteristic 
for the commonly known mixed state. We consider an alternative model for 
describing the mixed state for $\kappa \gg 1$ and magnetic fields close to 
$H_{c2}$. We argue that with  decreasing magnetic field the traditional 
vortex structure is adopted via a first order phase transition, revealed by 
discontinuities in the magnetization as well as the resistivity. 
\end{abstract}

\maketitle

It is commonly accepted that the mixed state of a type-II superconductor is 
characterized by the penetration of an external magnetic field into the 
sample along quantized vortex lines or vortices. The spatial extent of the 
superconducting order parameter $\psi (r)$ near a vortex axis is determined 
by circular currents flowing around the vortex line. The velocity of the 
relevant charge carriers, i.e., the Cooper pairs, diverges at the center of 
the vortex line (see, e.g., Ref. \cite{1}), thus reducing the 
superconducting order parameter progressively until it vanishes on the 
vortex axis. In most cases the vortices may thus be considered as thin 
normal filaments embedded in a superconducting environment. A completely 
different situation may, however, be established in superconductors with a 
Ginzburg-Landau (G-L) parameter $\kappa \gg 1$. If the applied magnetic 
field is close to the upper critical field $H_{c2}$, the distance between 
adjacent vortex cores is much smaller than the magnetic field penetration 
depth $\lambda (T)$. In this case, there is practically no expulsion of the 
magnetic field from superconducting regions and the density of shielding 
currents is negligibly small. These circumstances are naturally unfavorable 
for the formation of common Abrikosov vortices and hence the character of 
the mixed state in magnetic fields close to $H_{c2}$ may be very different 
from that adopted at lower fields. 

In this paper we consider that in magnetic fields close to $H_{c2}$, the 
natural alternative to the conventional vortex structure is the formation 
of superconducting filaments, embedded in the matrix of the normal metal. 
The properties of such superconducting filaments may be analyzed by 
numerically solving the G-L equations \cite{2}. First, we consider a single 
cylindrical superconducting filament in an infinitely extended normal metal. 
The magnetic field inside the normal metal is oriented parallel to the 
filament and its value  is set to $H_{0} < H_{c2}$. For the numerical 
analysis we have chosen cylindrical coordinates $(r,\phi,z)$ with the 
$z$-axis parallel to the filament and $r = 0$ at its center. The G-L 
equations for an infinitely long cylindrical filament 
\begin{equation}
-{1 \over {\kappa ^2}}\left( {{{d^2\psi } \over {dr^2}}+{1 \over r}{{d\psi} 
\over {dr}}} \right)+\left( {A^2-1} \right)\psi +\psi ^3=0
\end{equation}
and
\begin{equation}
{d \over {dr}}\left( {{{dA} \over {dr}}+{A \over r}} \right)=A\psi ^2.
\end{equation}
may thus be solved in one-dimension, with all quantities depending only 
on the radial coordinate  $r$. Here, $\psi$ is the superconducting order 
parameter normalized by its equilibrium value in a bulk superconductor; 
$A$ is the vector potential of the magnetic field expressed in units of 
$\sqrt{2}H_{c} \lambda (T)$, where $H_{c}$ is the thermodynamic critical 
field. The coordinate $r$ is measured in units of $\lambda (T)$.

The boundary conditions are set at the center of the filament ($r = 0$). 
For the superconducting order parameter $\psi (r)$, we have $d\psi /dr| 
_{r=0} = 0$ and $\psi (0) = const$. For the vector potential $A(r)$, we 
have $A(0) = 0$ and curl $A|_{r=0} = H(0)$, the value of the magnetic 
field at the center of the filament. We have checked that the solutions 
with $A(0) = 0$, equivalent to a vanishing supercurrent at $r = 0$, 
correspond to a minimum of the free energy. By integration of the G-L 
equations with these boundary conditions, we obtain $\psi (r)$, $A(r)$ and 
the distribution of the magnetic field $H(r)$ around the filament. We also 
calculated the Gibbs free energy $F_{fil}$ of the filament per unit length. 
For a single filament, $F_{fil}$ reaches its minimum if, at the boundary 
with the normal metal, the order parameter vanishes with a zero derivative, 
i.e., $\psi (r) \rightarrow 0$ and $d\psi /dr \rightarrow 0$ for $r 
\rightarrow \infty$.
\begin{figure}[ht]
 \begin{center}
  \epsfxsize=0.95\columnwidth \epsfbox {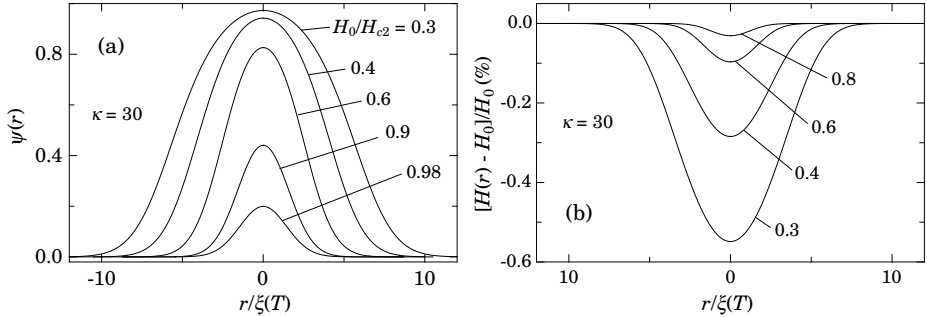}
  \caption{Radial variation of (a) the normalized order parameter 
           $\psi (r)$ and (b) the magnetic field for different 
		   values of $H_{0}/H_{c2}$. $\xi (T)=\lambda (T)/ \kappa$ 
		   is the G-L coherence length.}
 \end{center}
\end{figure}

The calculations have been made for 3 values of $\kappa =$ 10, 30, and 
100. Fig. 1(a) shows the calculated  profiles of the normalized 
superconducting order parameter $\psi (r)$ for one filament and, in Fig. 
1(b), corresponding profiles of the magnetic field have been plotted. For 
all $\kappa$ values between 10 and 100 and $H_{0}/H_{c2} \ge 0.3$, these 
profiles are independent of $\kappa$. In this approach, all characteristics 
of the filament are uniquely determined by the value of the applied magnetic 
field $H_{0}$. 

\begin{figure}[!h]
 \begin{center}
  \epsfxsize=0.95\columnwidth \epsfbox {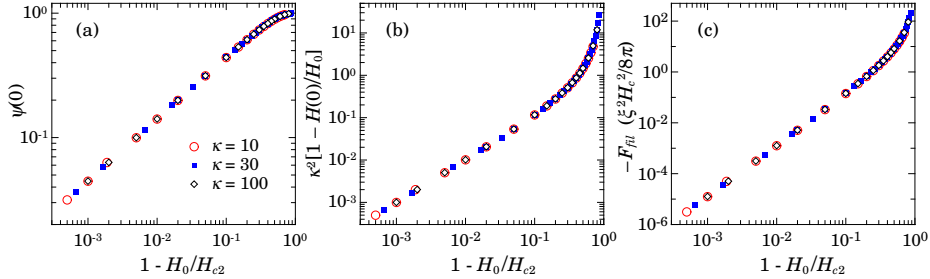}
  \caption{Different parameters of a single superconducting filament as 
           functions of $(1 - H_{0}/H_{c2})$. (a) The amplitude of the 
		   normalized superconducting order parameter at the center of 
		   the filament. (b) The normalized difference between the 
		   magnetic field at the center of the filament and $H_{0}$, 
		   multiplied by $\kappa^{2}$. (c) The Gibbs free energy of the 
		   filament $F_{fil}$ (per unit of length) in units of $\xi^{2}(T) 
		   H_{c}^{2}/8\pi$ with respect to that of the normal metal in a 
		   magnetic field equal to $H_{0}$.}
 \end{center}
\end{figure}

In Figs. 2(a-c) we display various quantities, plotted versus $(1 - 
H_{0}/H_{c2})$. Figure 2(a) shows the  amplitude of the normalized G-L 
order parameter at the center of the filament; in Fig. 2(b) we have 
plotted the product $\kappa^{2} [1 - H(0)/H_{0}]$, and in Fig. 2(c) we 
show the free energy of the filament $F_{fil}$. It may be seen that while 
$\psi (0)$ and $F_{fil}$ are independent of $\kappa$, the expulsion of 
the magnetic field from the center of the filament, $[1 - H(0)/H_{0}]$, 
is inversely proportional to $\kappa^{2}$, i.e., the magnetic moment of 
the filament rapidly decreases with increasing $\kappa$. According to Fig. 
2(c) the free energy of the filament with respect to that of the normal 
metal is negative and it vanishes at $H_{0} = H_{c2}$. This obvious but 
gratifying result demonstrates that the superconducting phase can only be 
stable in magnetic fields below $H_{c2}$.

So far, we have considered a single superconducting filament in a normal 
metal matrix and we now consider the interaction between the filaments. 
Because the current density in the normal metal between the filaments is 
zero there is no long-range interaction between them and the contribution 
to the free energy due to filament-filament interaction is zero unless the 
neighboring filaments are very close to each other. In order to show that 
the short-range interaction between the filaments is repulsive, we introduce 
the velocity of superconducting electrons $v_{s} = j_{s}/(2e \psi^{2})$, 
where $j_{s}$ is the supercurrent, $c$ is the speed of light, and $e$ is 
the electron charge. In our case $v_{s}$ increases with the distance from 
the axis of the filament approximately proportionally to $r$. The 
velocities $v_{s}$ arising from the neighboring filaments have opposite 
directions in the space between them. Because $v_{s}$ cannot have 
discontinuities in the superconducting phase, the filaments cannot merge, 
but must always be divided by a boundary where the order parameter $\psi 
\equiv 0$ \cite{4}. The situation in the boundary region is very similar to 
that arising in the center of the Abrikosov vortex where $v_{s}$ also 
changes its sign, demanding that $\psi = 0$ along the vortex axis. 
\begin{figure}[h]
 \begin{center}
  \epsfxsize=0.6\columnwidth \epsfbox {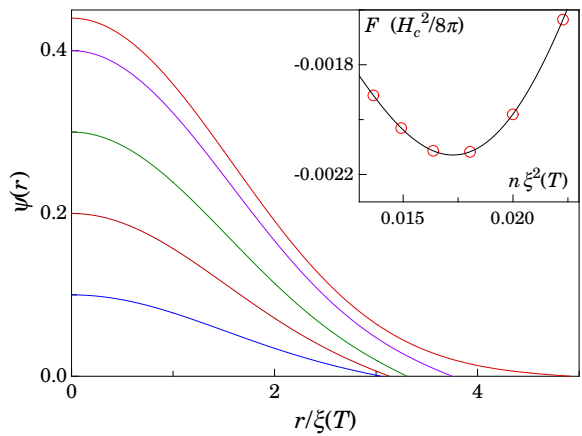}
  \caption{Profiles of the order parameter for different values of $\psi 
           (0)$ and $H_{0}/H_{c2}=0.9$. In this case, as well as for a 
           single filament, these profiles are independent of $\kappa$ 
           for $10< \kappa <100$. The inset shows the total free energy of 
           the system of filaments $F=n \tilde{F}_{fil}$ as a function of 
		   their density. The solid line is a guide to the eye.}
 \end{center}
\end{figure}

Because the number of  filaments in the sample is only limited by a 
short-range repulsion, they are expected to form a rather dense triangular 
configuration as is illustrated in the inset to Fig 4(b). In order to 
analyze the resulting profile of an individual filament, we again use Eqs. 
(1) and (2) but, in order to satisfy the condition $\psi = 0$ between the 
filaments, the boundary condition $d\psi /dr|_{r \rightarrow \infty} = 0$ 
is abandoned. Several solutions of the G-L equations are shown in Fig. 3. 
In this case, the order parameter vanishes with a non-zero derivative and 
we may introduce the radius of the filament as the value of $r=D/2$ at the 
point where $\psi =0$. 

As may be seen in Fig. 3, the diameter $D$ of the filament decreases with 
decreasing value of $\psi (0)$. The important consequence of the G-L 
equations is that the diameter $D$ of the filament cannot be smaller than 
a certain minimal value $D_{\min}$. The only solution of the G-L equations 
for $D \le D_{\min}$ is $\psi (r) \equiv 0$. The dependence of $D_{\min}$ 
on $(1-H_{0}/H_{c2})$ is shown in Fig. 4a. We note that also for 
superconducting lamellae a minimal thickness $\delta _{\min}$ exists. The 
dependence of $\delta _{\min}$ on the applied magnetic field is shown in 
Fig. 4a, as well. Both $D_{\min}$ and $\delta _{\min}$ decrease with 
decreasing magnetic field. It has been shown that $\delta _{\min}|_{H_{0}=0} 
= \pi \xi (T)$ \cite{5}. The corresponding value of $D_{\min}|_{H_{0}=0} 
\approx 4.81 \xi (T)$.
\begin{figure}[ht]
 \begin{center}
  \epsfxsize=0.9\columnwidth \epsfbox {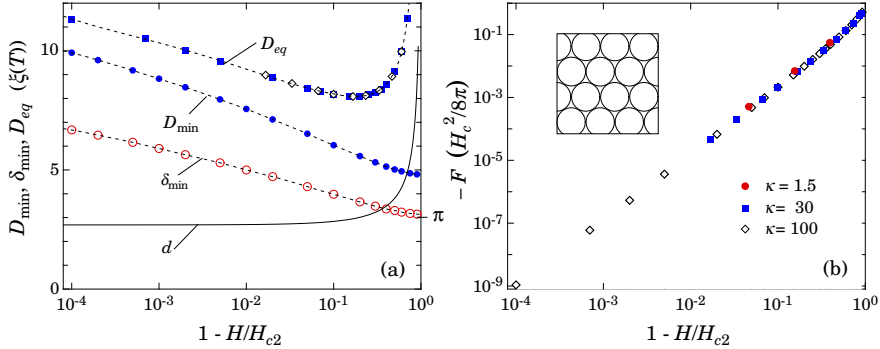}
  \caption{(a) The minimal thickness of superconducting lamellae, 
           $\delta_{\min}$, the minimal diameter of the superconducting 
		   filament, $D_{\min}$, and the equilibrium value $D_{eq}$ versus 
			$(1-H_{0}/H_{c2})$. The period $d$ of the vortex structure is 
			shown as the solid line. The dashed lines are guides to the 
			eye. The Gibbs free energy $F$ of the system of filaments (per 
			unit of volume) with respect to that of the normal metal in a 
			magnetic field equal to $H_{0}$. The inset represents the 
			triangular lattice of superconducting filaments}
 \end{center}
\end{figure}

In the following we assume that the distance between the filaments is equal 
to their diameter $D$ as is shown in the inset to Fig. 4(b). For a 
triangular lattice, the density $n$ of filaments is thus $n=2/(\sqrt 3 
D^{2})$. In order to evaluate the equilibrium density of filaments, we plot 
the total free energy of the system of filaments $F=n \tilde{F}_{fil}$, 
where $\tilde{F}$ is the Gibbs free energy calculated for solutions of the 
type shown in Fig. 3, versus $n$ as is illustrated in the inset to Fig. 3. 
The position of the minimum on this curve corresponds to the equilibrium 
value of $n$. The magnetic field dependence of the equilibrium diameter 
$D_{eq}$ of the filaments is presented in Fig. 4(a). As may be seen, 
$D_{eq}$, which is the period of the triangular lattice of the filaments, 
is considerably larger than the equivalent quantity $d$ for the 
vortex structure. 

The dependencies of $\psi (0)$ and $(1-H(0)/H_{0})$ on the magnetic field 
for an equilibrium filament configuration practically coincide with those 
calculated in the single-filament approximation for $H_{0} /H_{c2} \ge 
0.5$ (see Figs. 2(a) and 2(b)).

Fig. 4(b) shows the free energy of the system of filaments as a function 
of $(1-H_{0}/H_{c2})$. The energy was calculated assuming that the 
filaments are cylindrical. We have to adopt this approach, in order to use 
one-dimensional G-L equations which are the basis of our consideration. 
It is obvious, however, that, because of their mutual interaction,the 
filaments should adopt a hexagonal rather than a circular cross-section. 
We note, however, that this simplification may only result in free energy 
values that are slightly higher than the actual ones.
\begin{figure}[t]
 \begin{center}
  \epsfxsize=0.6\columnwidth \epsfbox {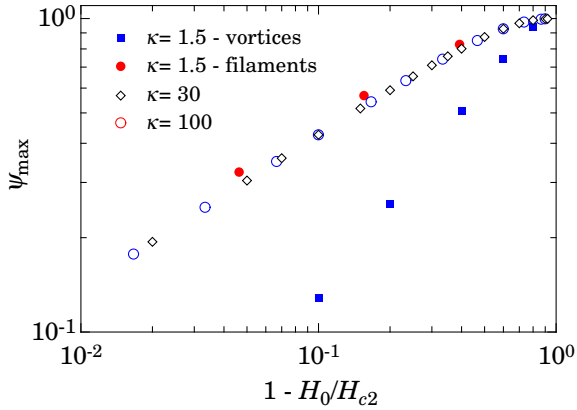}
  \caption{The maximum amplitude of the order parameter for the equilibrium 
           configuration of filaments and for the vortex lattice. The data 
		   for vortices is taken from Ref. \protect \cite{5B}.}
 \end{center}
\end{figure}

As we have argued, the system of quantized vortices is not the only way to 
realize the mixed state of type-II superconductors which may also be 
established by an ensemble of superconducting filaments. At low magnetic 
fields, where the equilibrium diameter of filaments $D_{eq} \ge \lambda 
(T)$, the filaments are unstable against the formation of a vortex line in 
the center of the filament. In order to identify the stable arrangement in 
higher fields, the free energies for these two configurations should be 
compared. To our knowledge the only study in which the two-dimensional G-L 
equations have been solved for the vortex lattice is presented in Ref.  
\cite{5B}. Fig. 5 shows the results of our calculations of the order 
parameter amplitude $\psi _{\max}$ versus $(1-H_{0}/H_{c2})$ for filaments, 
together with the results of Ref. \cite{5B}. In the high-magnetic-field 
limit, the term proportional to the magnetic moment in the Gibbs free 
energy may be neglected and the density of the free energy reduces to $F 
\approx - \psi ^4(r) H_{c}^{2}/8 \pi$. As may clearly be seen in Fig. 5, 
the amplitude of $\psi$ for the vortex lattice is considerably smaller than 
that for the filaments and the ratio $\psi _{ \max} 
(\textnormal{filaments})/\psi _{ \max} (\textnormal{vortices})$ increases 
with increasing magnetic field. Thus, in sufficiently high magnetic fields, 
the free energy for the configuration of superconducting filaments is 
expected to be lower than that for the vortex lattice.

The properties of a mixed state consisting of superconducting filaments 
are quite different from those of the Abrikosov vortices. First, because 
the filaments are always separated by normal conducting regions, the sample
resistance for currents perpendicular to the direction of the magnetic 
field never vanishes and the true zero-resistance superconducting state may 
be achieved only after the transition to the vortex structure. Second, in 
the case of filaments, the magnetic flux faces no barriers to move in or 
out of the sample and the magnetization of the sample must be reversible, 
independent of whether the filaments are pinned or not.

The analysis presented in this paper shows that, in magnetic fields close 
to $H_{c2}$, the mixed state may well consist of a triangular lattice of 
superconducting filaments separated by regions where the superconducting 
order parameter $\psi =0$. With decreasing external magnetic field, the 
configuration of superconducting filaments necessarily has to undergo a 
transition to the conventional mixed state, involving Abrikosov vortices. 
The value of the transition field is determined by the free-energy balance 
between these two configurations which cannot be determined without more 
precise calculations of the free energy for both cases. The transition from 
one type of mixed state to the other involves a complete change of topology 
and must be accompanied by discontinuities in both the resistivity and the 
magnetic moment of the sample. We also expect some hysteresis, as well as a 
latent heat, dictated by the discontinuity of the magnetization. In other 
words, this transition is expected to exhibit all the features of a 
first-order phase transition. Such transitions are observed in high 
temperature superconductors at $H < H_{c2}$ and are usually attributed to 
the melting of the vortex lattice \cite{9,10,11,12}.

\verb" "

We wish to thank R. Monnier for numerous stimulating discussions.


\begin{thebibliography}{10}
\bibitem {2} The calculation procedure is described in Ref. 3. The only 
difference is that here we use the Ginzburg-Landau equations written in 
cylindrical coordinates.
\bibitem {4} The fact that superconducting domains in the mixed state of 
type-II superconductor cannot merge into one, may even more clearly be 
demonstrated if we consider two superconducting planes instead of 
filaments. We choose the vector-potential $A=0$ along the boundary 
between the planes and the order parameter in its complex form. In this 
case $v_{s}$ is proportional to $\nabla \phi$, where  $\phi$ is the phase 
of the order parameter. Because $\nabla \phi$ changes its sign if we go 
from one superconducting plane to the other, it is clear that the order 
parameters of these two planes cannot be forced to merge. Superconducting 
planes, however, are unstable against the formation of a system of 
Abrikosov vortices between them. The distance $L$ between the vortices 
may be found from the condition $L \nabla \phi = \pi$. In the case of 
superconducting filaments, such boundary vortices cannot be formed because 
the phase difference $\Delta \phi$ along the filament-filament boundary is 
much less than $\pi$.
\end{thebibliography}
\end{document}